\documentclass[10pt,letterpaper,twocolumn]{article} 

\usepackage{ol2}
\usepackage[draft]{hyperref}
\usepackage{amsmath}
\usepackage{graphicx}
\begin{document}
\newcommand{\be}{\begin{equation}}
\newcommand{\ee}{\end{equation}}
\twocolumn[ 

\title{All-optical Realization of Quantum Ratchets}

\author{Clinton Thompson,$^{1}$ Gautam Vemuri,$^{1,*}$ and G.S. Agarwal$^{2}$}

\address{
$^1$Department of Physics, Indiana University Purdue University Indianapolis (IUPUI), Indianapolis, IN 46202-3273
\\
$^2$Department of Physics, Oklahoma State University, Stillwater, OK 74078\\
$^*$Corresponding author: gvemuri@iupui.edu
}

\begin{abstract}A theoretical realization of an all-optical quantum ratchet is proposed in a medium composed of an array of coupled waveguides.  By coupling light into two adjacent waveguides, and calculating the expectation value for the position space operator, we demonstrate the quantum ratchet like behavior of this optical system. \end{abstract}


 ] 

\noindent A ratchet is a device in which there is directed motion of a particle in one direction, but in which motion is blocked in the opposite direction.  The Brownian motion ratchet has attracted considerable attention since the ratchet paradox was resolved by Feynman\cite{Feynman1963}. The classical ratchet, which has been studied extensively in the context of biological motors, requires a combination of an asymmetric potential and a dissipative, random noise process, which together are responsible for the directed ratchet motion \cite{Austumian2002}. In recent years, a quantum version of the ratchet has been demonstrated in Bose-Einstein condensates.  Unlike the classical ratchet which requires a symmetric input and an asymmetric potential, the quantum ratchet does not require a dissipative process, and one can utilize a symmetric potential with an asymmetric input.  The directed motion is a result of an underlying asymmetry between the potential in which the particle is moving and the quantum mechanical density distribution. This is illustrated in Fig. 1.  In part (a), there is a symmetry between the potential and the density distribution, and there is no net motion of the particles because when the potential is ON, the particle is always trapped at the minima of the potential. In part (b), the asymmetry between the potential and the density distribution leads to net motion of the particle.
\begin{figure}[ht]
  \centering
    \includegraphics[width=0.4\textwidth]{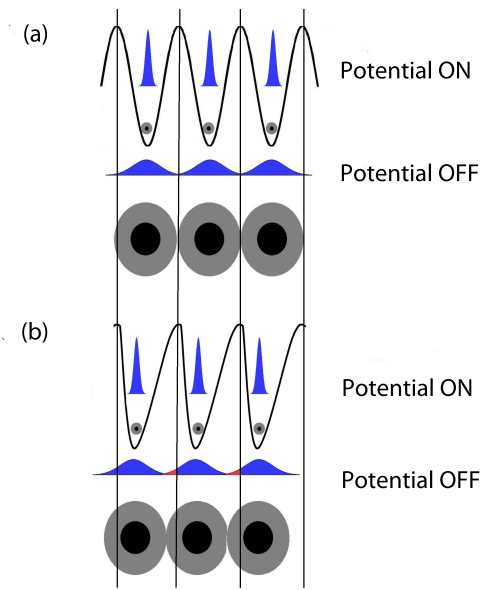}
\caption{(Color Online) A rachet results from the asymmetry of the potential and initial density distribution. If the potential (black) and the distribution (blue) are symmetric (a), there is no net motion of the particles (gray and black circles). However, if the potential is not symmetrical with respect to the distribution (b), there is a net flow of particles to the left; following \cite{Ramareddy2008}.}
  \label{Figure1}
\end{figure}
A standard model to describe quantum ratchets is the kicked rotor given by the Hamiltonian,
\be 
H = \frac{p^{2}}{2m}+\hbar V_{0}\cos(2kx-\theta)\sum_{n}\delta(t-nT)
\ee
where the periodic potential is produced by a standing laser field. This field is applied at $t = nT$ where $n$ is an integer and the field consists of a series of short pulses. The initial state, $\psi$, of the particle is a combination of states with different momenta,
\be 
\psi(x)=\psi_{0}(e^{iqx}+\alpha e^{i\varphi}e^{iqx+2ikx})
\ee 

The phase $\varphi$ and the parameter $\alpha$ introduce the asymmetry between the potential and the density distribution and lead to directed motion of the particles. This has been extensively studied both theoretically and experimentally \cite{Sadgrove2007}. In what follows, we suggest an all-optical realization of quantum ratchets by using coupled array of waveguides. A previous, related work, has investigated the optical ratchet effect in coupled waveguide arrays \cite{Longhi2009}. We note that such arrays of waveguides have proved to be of great utility in the realization of a number of effects from condensed matter physics and quantum physics, such as Bloch oscillations \cite{Peschel1998}, Anderson localization \cite{Lahini2008,Thompson2010}, quantum Zeno effect \cite{Biagaioni2008}, electromagnetically induced transparency \cite{Longhi2010b}, Dirac zitterbewegung \cite{Longhi2010,Longhi2010a}, Talbot effect \cite{Iwanow2005}, quantum random walks \cite{Perets2008}, surface solitons \cite{Suntsov2006,Smirnov2006}, Zener tunneling \cite{Trompeter2006a}, and a quantum bouncing ball \cite{Longhi2008}, among others.

Consider an array of waveguides, shown in Fig. 2, in which the neighboring waveguides are evanescently coupled.  This system is described by a tight binding Hamiltonian of the form,
\be 
H = \hbar  \sum_{j} \beta(j)a^{\dagger}_{j}a_{j}+\hbar C\sum_{j}(a^{\dagger}_{j+1}a_{j}+a^{\dagger}_{j-1}a_{j})
\ee
where the array is labeled by the index j and $\beta(j)$ is related to the refractive index of the jth waveguide.  In Eq. (3), $C$ is the coupling between adjacent waveguides, and $a_{j}^{\dagger} \left( a_{j} \right)$ is the creation (annihilation) operator for the jth waveguide. These obey the Heisenberg equations,
\be 
\dot{a}_{j}=-i\beta(j)a_{j}-iC(a_{j+1}+a_{j-1})
\ee
From now onwards, we consider a classical description (for quantum mechanical descriptions, see \cite{Reimann1997}). Thus, the $a_{j}$'s would be regarded as numbers.

We now write
\be 
a_{j}=\frac{1}{\sqrt{2\pi}}\int^{\pi}_{-\pi}\tilde{a}(k)e^{ikj}dk
\ee
\be 
\tilde{a}_{j}=\frac{1}{\sqrt{2\pi}} \sum_{j} a_{j}e^{-ikj}
\ee 
We can then write the equations in the Fourier space using Eq.(4) and Eq.(5), as
\be 
\dot{\tilde{a}}(k)=-2iC\tilde{a}(k)\cos k-i\beta\left(\frac{\partial}{\partial(-ik)}\right) \tilde{a}(k)
\ee 
This should now be compared with the Hamiltonian in Eq. (1). In Eq. (7), we already have the periodic potential. In Eq. (1), we have a term that is quadratic in $p$, i.e. a term like $\frac{\partial^{2}}{\partial x^{2}}\,$. However, for a Bose-Einstein condensate (BEC) \cite{Dana2008}, the distribution in $p$ is very narrow and thus an expansion around some mean momentum would reduce $p^{\,2}$ to a term linear in $p$. On the other hand, for the waveguide structure we can make the choice,
\be 
\beta(j)=j\beta
\ee
and then Eq.(7) reduces to 
\be 
\dot{\tilde{a}}(k)=-2iC\tilde{a}(k)\cos k-ij\beta\frac{\partial}{\partial(-ik)} \tilde{a}(k)
\ee 

\begin{figure}[ht]
  \centering
    \includegraphics[width=0.6\textwidth]{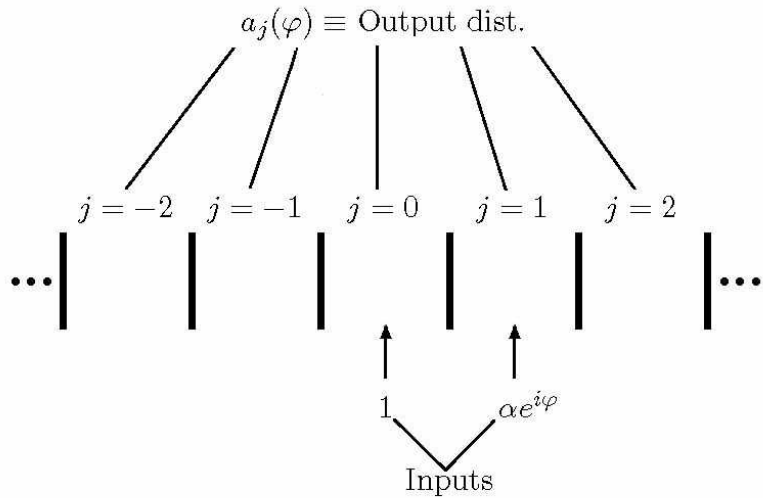}
\caption{A large waveguide array with the input fields shown.}
\label{waveguide}
\end{figure}

The choice of the waveguide index, Eq. (8), has been experimentally realized in studies on Bloch oscillations in waveguide structures. It should be borne in mind that the roles of position and momentum have been reversed in the optical realization of quantum ratchets. In Eq. (1), the periodic potential is in coordinate space whereas for the waveguide structure the periodic potential is in the Fourier space, $k$. In the quantum case, one studies the mean values of $p$, i.e. $<p>$, to obtain directed motion. In our waveguide system, we have to study the behavior in the �site� space, $j$. In the quantum case, we specify the wavefunction, Eq. (2), in the coordinate space. Here, the corresponding analog would be the value of $\tilde{a}(k)$ in the $k$ space, which can be translated to the site space $j$. The appropriate input conditions on the field amplitude will be 
\be 
a_{j}(t=0)=\delta_{j,0}+\alpha \delta_{j,1}e^{i\varphi}
\ee 
We assume that the array index is from $j = -\infty$ to $\infty$, see Fig.~\ref{waveguide}, where the $j = 0$ is the middle waveguide. The $\vert a_{j}\vert^{2}$ play the role of the density distribution. The analogs of quantum ratchets that we would find would be due to $\alpha  \ne 0$, $\varphi \ne 0$. The familiar Bloch oscillations occur for $\alpha=0$. The possibility of additional interference effects due to $\alpha \ne 0$ leads to the realization of the analog of quantum optical ratchets.

In order to study the ratchet problem, we calculate the expectation values for $<j>$ and $<j^{2}>$ which are analogous to the expectation values for momentum and energy respectively. These expectation values are calculated as $<j>=\sum_{j=-\infty}^{\infty} j I_{j}$ and  
$<j^{2}>=\sum_{j=-\infty}^{\infty} j^{2} I_{j}$ where $I_{j}$ is the output intensity of the $j^{th}$ waveguide.
We write the solution to the coupled equations, Eq. (4), in the form 
\be 
a_{j}=\sum_{j'} G_{j,j'} a_{j'}(0)
\ee
where $a_{j}(0)$ is given by Eq. (10) and the Green's function is given by Eq. (21) from \cite{Rai2009}
\begin{align} 
{G_{j,j'}}&=\exp\left[i\beta z+\frac{i(j-j')(\beta z-\pi)}{2}\right]\nonumber\\
&\quad\times J_{j'-j}\left[ \frac{4C}{\beta}\sin\left( \frac{\beta z}{2}\right)\right].
\end{align}
Using the initial condition  of Eq. (10), the output intensity from the waveguide array can be written as 
\begin{align}
{I_{j}} &=\vert G_{j,0} \vert^{2}+\vert \alpha G_{j,1} \vert^{2}+\alpha G_{j,0} G_{j,1}^{\ast}e^{i\varphi}+\alpha G_{j,0}^{\ast} G_{j,1}e^{i\varphi}\nonumber\\
&=\vert J_{-j}\left[ \frac{4C}{\beta}\sin\left( \frac{\beta z}{2}\right)\right] \vert^{2} +\vert \alpha J_{1-j}\left[ \frac{4C}{\beta}\sin\left( \frac{\beta z}{2}\right)\right]\vert^{2}\nonumber\\
&\quad-2\alpha J_{-j}\left[ \frac{4C}{\beta}\sin\left( \frac{\beta z}{2}\right)\right]J_{1-j}\left[ \frac{4C}{\beta}\sin\left( \frac{\beta z}{2}\right)\right]\nonumber\\
&\quad\times\sin\left(\frac{\beta z}{2}-\varphi\right).
\end{align}

Fig. 3(a) shows the output intensity from the waveguide array, given by Eq. (13), when $\alpha=0$. We note immediately from Fig. 3(b) that there is an asymmetry in the profile along the j-axis, and this asymmetry is evidence for directed motion. The case when $\alpha=0$, in Fig. 3(a), results in symmetric profiles, and corresponds to the well-known Bloch oscillations\cite{Rai2009,Peschel1998}.
\begin{figure}[ht]
  \centerline{
    \mbox{\includegraphics[width=0.225\textwidth]{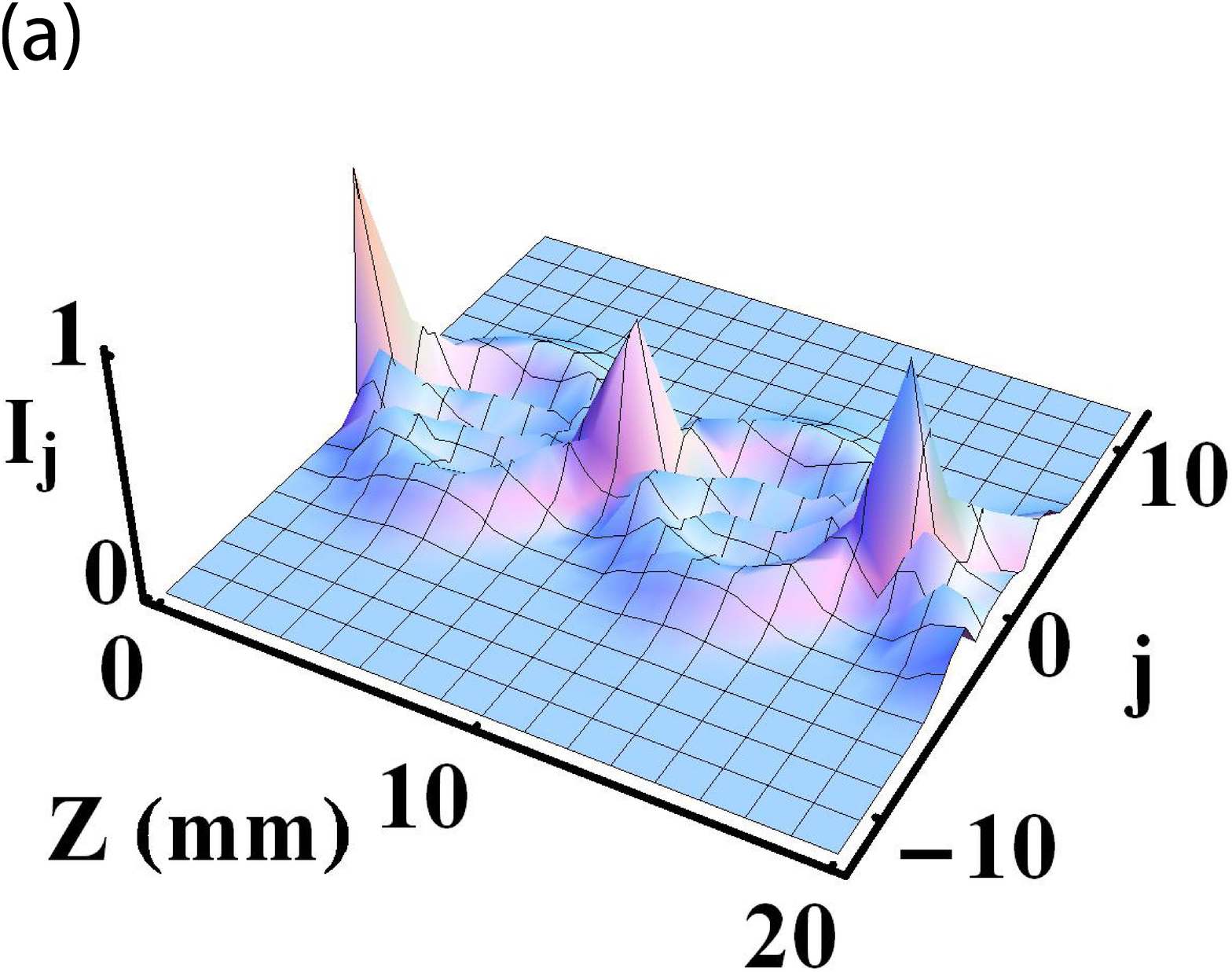}}
    \mbox{\includegraphics[width=0.225\textwidth]{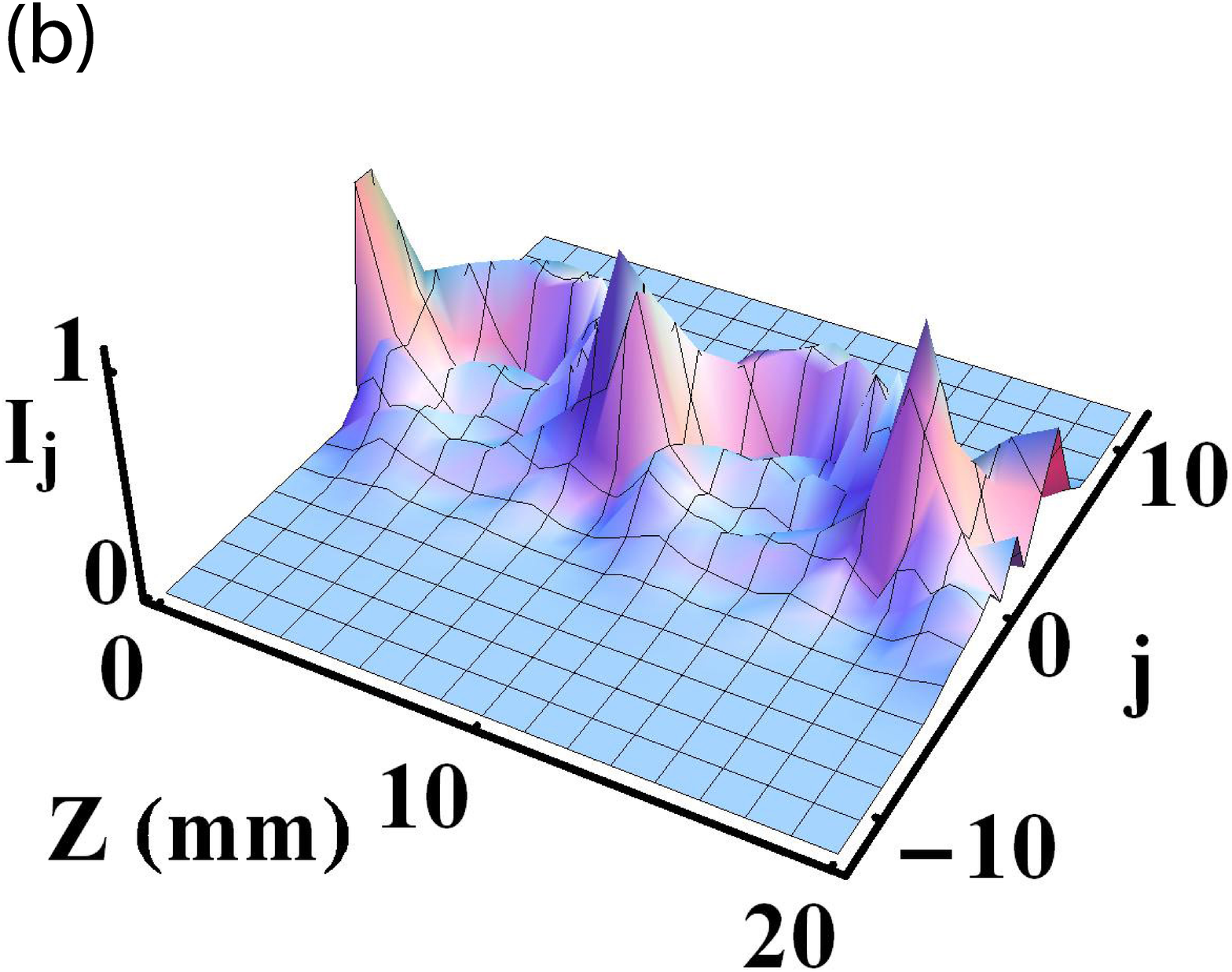}}
  }
  \caption{(Color Online) Output intensity distribution, $I_{j}$, as a function of $z$ for $j=-12,...,12.$ The values of the parameters are $\beta /C =0.73,$ $\varphi = 37^{\circ},$ and (a) $\alpha = 0,$ (b) $\alpha = 1.$}
\end{figure}

Using Eq. (13), the expectation value for the site position can be written as (after using properties of Bessel functions)
\be 
\sum_{j=-\infty}^{\infty}jI_{j}=\vert\alpha\vert^{2}+\frac{4\alpha C}{\beta}\sin\left(\frac{\beta z}{2}\right)\sin\left(\frac{\beta z}{2}-\varphi\right).
\ee
Similarly, the expectation value for the analog of energy is given by
\begin{align}
{\sum_{j=-\infty}^{\infty}j^{2}I_{j}} &=\vert\alpha\vert^{2}+\frac{4\alpha C}{\beta}\sin\left(\frac{\beta z}{2}\right)\sin\left(\frac{\beta z}{2}-\varphi\right)\nonumber\\
&\quad+\frac{1+\vert\alpha\vert^{2}}{2}\left(\frac{4 C}{\beta}\sin\left(\frac{\beta z}{2}\right)\right)^{2}.
\end{align}

For small values of z, the average position, Eq. (14), is proportional to $-\alpha \sin\left(\varphi\right) z,$ which shows that direction of transport for the photons is dependent on the the relative phase. Fig. 4(a) shows that the photons with a phase of $37^{\circ}$ are directed to the left in Fig. 2 for small values of $z$ while Fig. 5(a) shows the photons with a phase of $217^{\circ}$ are directed to the right. The directed motion is a result of the interference of the two photons as the second term in Eq. (14) is linear in $\alpha$. The analog of the energy is also linear in $z$ for small values of $z$ as a result of the interference since the second term in both Eq. (14) and(15) are identical. Also, the dependence of energy on the phase is identical to that of the momentum as seen in Fig. 4(b) and 5(b).   

\begin{figure}[ht]
  \centerline{
    \mbox{\includegraphics[width=0.2\textwidth]{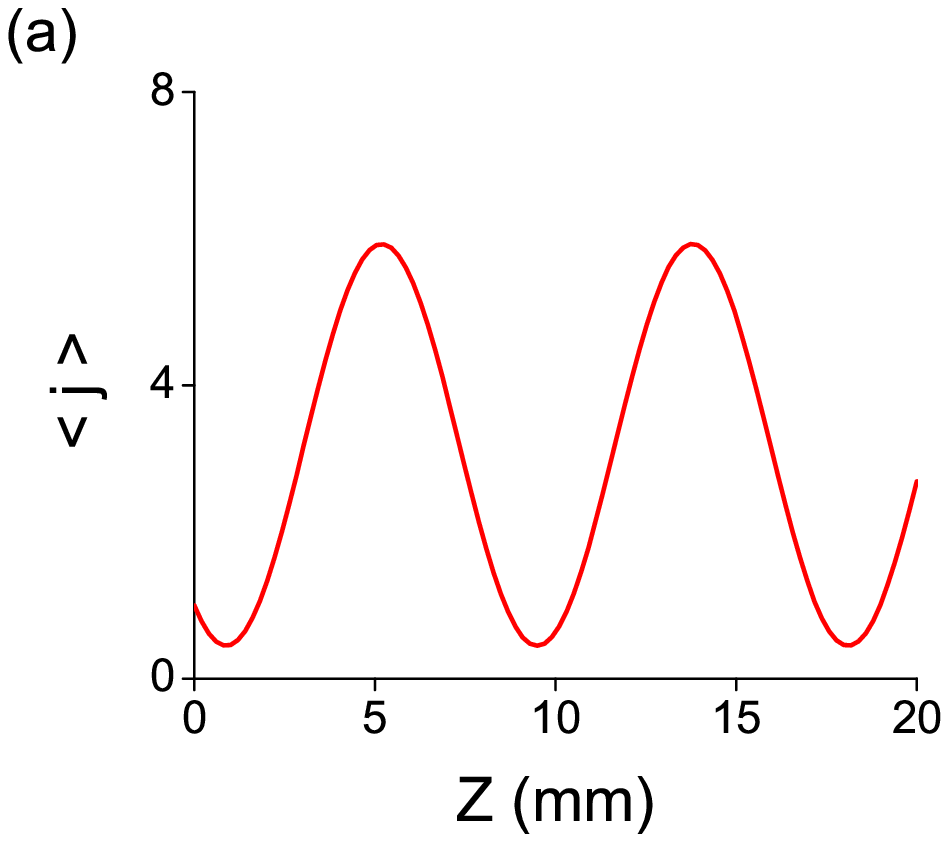}}
    \mbox{\includegraphics[width=0.2\textwidth]{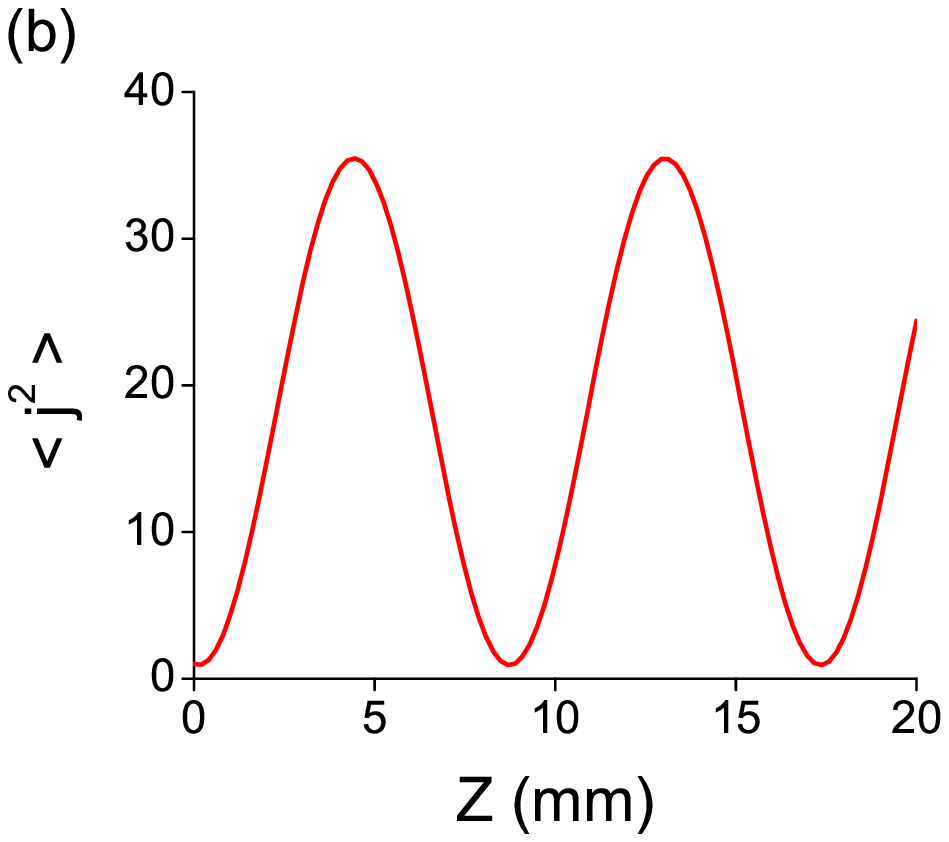}}
  }
  \caption{(Color Online) Plots showing ratchet like behavior for (a) the momentum and (b) the energy. The values of the parameters are $\alpha=1$, $\beta /C =0.73,$ and $\varphi = 37^{\circ}$.   }
  \label{alpha1}
  \end{figure}

\begin{figure}[ht]
  \centerline{
    \mbox{\includegraphics[width=0.2\textwidth]{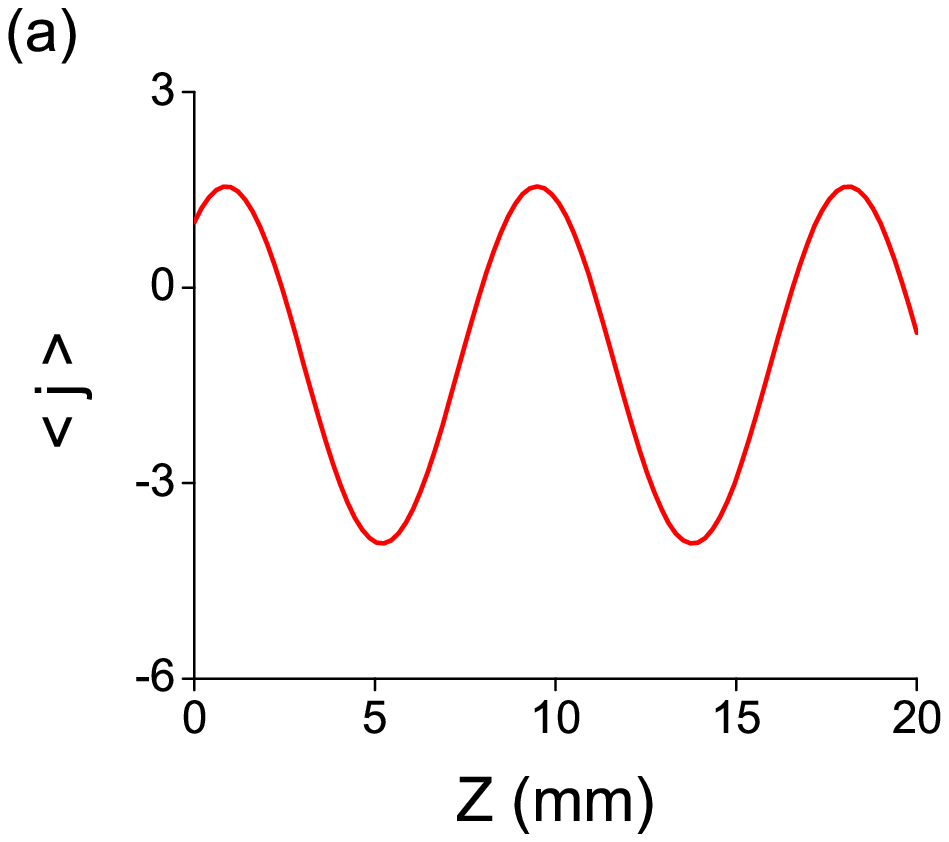}}
    \mbox{\includegraphics[width=0.2\textwidth]{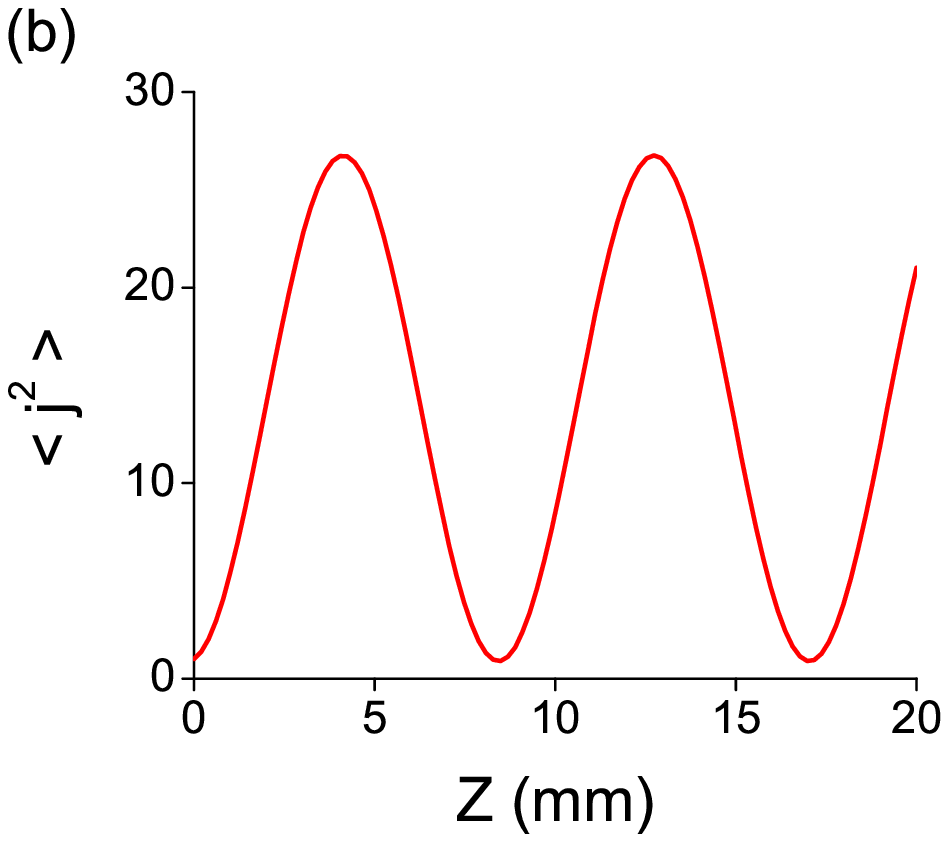}}
  }
  \caption{(Colore Online) Plots showing ratchet like behavior for (a) the momentum and (b) the energy. The values of the parameters are $\alpha=1$, $\beta /C =0.73,$ and $\varphi = 217^{\circ}$.   }
  \label{alpha53}
  \end{figure}
    
In summary, we have described an all-optical realization of the quantum ratchet in a system of evanescently coupled waveguide arrays.  Starting from the tight-binding Hamiltonian that is used to describe such a system, one can derive an equation of motion that is analagous to the kicked rotor model of the quantum ratchet. The principal difference between the theoretical frameworks is that in our optical realization of the quantum ratchet, the periodic potential is in the Fourier space instead of coordinate space.  This requires one to study the optical ratchet behavior in the site-space. This is fortunate because all waveguide studies are done in site-space.  We have demonstrated that by choosing a suitable input to the waveguide array that consists of a linear combination as described by Eq. (10), one can recover the Bloch oscillations as a limiting case of the ratchet behavior when $\alpha=0$.

C.T. was supported by a GAANN award from the US Department of Education to G.V.

\end{document}